\documentclass[aps,pra,superscriptaddress,amsmath,amssymb,preprintnumbers,showpacs]{revtex4}
\usepackage{amssymb}
\usepackage{graphicx}
\usepackage{bm}
\usepackage{empheq}
\usepackage{color}
\usepackage{upgreek}
\usepackage{cancel}
\usepackage{slashed}

\newcommand{\Ignore}[1]{}

\newcommand{\Ket}[1]{\left\vert #1\right\rangle}
\newcommand{\Bra}[1]{\left\langle #1\right\vert}

\renewcommand{\eqref}[1]{Eq.~(\ref{#1})}

\newcommand{\dom}{{\cal D}}

\def\e{\mathrm{e}}
\def\ii{\mathrm{i}}

\begin{document}

\title{Resonant transitions due to changing boundaries}

\author{Fabio Anz\`{a}}
\address{Clarendon Laboratory, University of Oxford, Parks Road, Oxford OX1 3PU, United Kingdom}
\author{Antonino Messina}
\address{Dipartimento di Matematica, Universit\`a di Palermo, Via Archirafi 36, I-90123 Palermo, Italy}
\author{Benedetto Militello}
\address{Dipartimento di Fisica e Chimica, Universit\`a di Palermo, Via Archirafi 36, I-90123 Palermo, Italy}

\begin{abstract}

The problem of a particle confined in a box with moving walls is studied, focusing on the case of small perturbations which do not alter the shape of the boundary (\lq pantography\rq). The presence of resonant transitions involving the natural transition frequencies of the system and the Fourier transform of the velocity of the walls of the box is brought to the light. The special case of a pantographic change of a circular box is analyzed in depth, also bringing to light the fact that the movement of the boundary cannot affect the angular momentum of the particle.

\end{abstract}

\pacs{03.65.-w, 03.65.Db}

\maketitle

\section{Introduction}\label{sec:Introduction}

Resolution of Schr\"odinger equations with time-dependent Hamiltonian operators is a challenging task. In fact, exact resolutions are rare and limited to specific classes of problems~\cite{ref:Barnes2012,ref:Messina2014,ref:Simeonov2014, ref:Sinitsyn2018}. On the contrary, in most of the cases one can solve the dynamical problem only under special assumptions and with some approximations, as it happens in the presence of weak interactions which legitimate the use of a perturbative approach~\cite{ref:Aniello2005,ref:Rigolin2008,ref:Zagury2010}, or when specific commutation relations are satisfied~\cite{ref:commuting1, ref:commuting2, ref:Militello2015a}.  An important class of time-dependent hamiltonians is that of slowly changing ones, since they lead to adiabatic evolutions~\cite{ref:Messiah}, with very important applications in quantum system manipulation spanning from Landau-Zener model~\cite{ref:LZ-history-1,ref:LZ-history-2,ref:LZ-history-3,ref:LZ-history-4} and its generalizations~\cite{ref:Fishman1990, ref:Vitanov1999, ref:Shytov2004, ref:Militello2015b, ref:Dodonov2016} to STIRAP protocols~\cite{ref:STIRAP_rev1, ref:STIRAP_rev2, ref:STIRAP_rev3, ref:Schiemann1993, ref:Vitanov1997, ref:Scala2011a, ref:Scala2011b} to the \lq fast counterpart\rq\, of adiabatic evolutions, i.e., the shortcuts to adiabaticity~\cite{ref:Berry2009, ref:Campo2012, ref:Campo2013}.
Another intriguing class of problems is given by periodic Hamiltonians, which allow for exploitation of some special recipes based on the Floquet theory~\cite{ref:Traversa2013,ref:Moskalets2002,ref:Shirley1965}.

In the panorama of systems with time-dependent Hamiltonians, a very special class is that of systems whose boundary conditions are time-dependent. Such kind of problems have been widely studied over the last decades in connection with the Casimir effect~\cite{ref:CasimirReview,ref:Wilson2010}, but in the past it has also been considered in connection with quantum mechanical  problems, from the Fermi quantum bouncer~\cite{ref:Fermi1949} to several works analyzing a \lq\lq free\rq\rq\, quantum particle in a box with moving walls~\cite{ref:Doescher, ref:Pinder, ref:Schlitt, ref:Dodonov}. The interest in such a class of problems is not only academic, and the relevance of moving boundaries has been discussed in cavity quantum electrodynamics~\cite{ref:Garraway2008} and in the physical scenario of trapped particles to propose new strategies for cooling atoms~\cite{ref:XiChen2009}. Recently, the role of the distortion of a trapping bounding potential has been studied and proved to be a crucial aspect in some chemical processes~\cite{ref:Dahms2017}.

Several other works appeared, dealing with specific box shapes~\cite{ref:Mousavi2012, ref:Mousavi2013}, aimed at giving  proper mathematical treatment of the Schr\"odinger problem in the presence of moving boundaries~\cite{ref:DiMartino}, and exploring the raise of correlations between different particles confined in the same time-dependent potential~\cite{ref:Mousavi2014}. Very recently, some works reporting on the numerical resolution of the dynamics of a particle confined in a one-dimensional box with moving walls has appeared~\cite{ref:Carrasco2017, ref:Duffin2018}.

In this paper, we analyze the dynamics of a quantum particle confined into a box whose walls are moving, and in particular we will focus on the possibility of exploiting moving boundaries to resonantly stimulate such particle. 

In Ref.~\cite{ref:Anza2015} we have shown that the problem of a particle confined in a two-dimensional (or three-dimensional) box with moving boundaries can be traced back to the problem of a particle with a varying mass, confined within a static box and subjected to a time-dependent potential. Following such an approach, we recast the original problem into the problem of a particle in a static box governed by a time-dependent effective Hamiltonian. It is intuitive and will be proven that if the walls move periodically, the effective time-dependent Hamiltonian will be periodic too, and with the same frequency. Through the exploitation of a time-dependent perturbative approach, we single out the presence of specific resonant transitions which are immediately traceable back to the analysis of the Fourier transform of the velocity of the boundary compared with the natural transition frequencies of the physical system. We will focus on two-dimensional problems but our results are immediately extendable to the three-dimensional case, and of course are essentially valid also for one-dimensional systems.

The paper is organized as follows. In the next section we introduce the physical problem, the relevant mathematical formulation and the proper unitary transformation which removes the boundary motion and make it \lq replaced\rq\, by a time-dependent term in the Hamiltonian. In the subsequent section we introduce the perturbative treatment singling out the presence of resonances and discussing the role of the Fourier transform of the boundary motion. In section \ref{sec:OscillatingCircle} we specialize the previous results to the case of a two-dimensional circular box. Finally, in section \ref{sec:Conclusion} we give some conclusive remarks.

\section{The Physical System and its Hamiltonian}\label{sec:PhysicalSystem}

\subsection{Two-dimensional box}

Let us consider a particle confined in a two-dimensional box whose contour, expressed in polar coordinates, is given by
\begin{equation}\label{eq:OriginalBoundary}
  r = \lambda(t)\gamma(\theta)\,,\qquad \theta \in [0, 2\pi]\,,
\end{equation}
where $\lambda(t)$ is a (dimensionless) smooth function of time, and $\gamma(\theta)$ is a smooth function of the angle $\theta$ such that $\gamma(2\pi)=\gamma(0)$ (which guarantees that the box is not open and the particle is confined). Because of the presence of $\lambda(t)$ the dimensions of the box change, but since the dependence of the radial coordinate of the box on $\theta$ and $t$ is factorized as $\lambda(t)\gamma(\theta)$ the shape of the box does not change. This is a kind of boundary modification that we call \lq pantographic\rq, and that corresponds to the fact that each point of the boundary moves only along the relevant \lq radial\rq\, direction.

The Hamiltonian describing the particle is simply given by the kinetic term:
\begin{subequations}
\begin{eqnarray}
  H &=& \frac{p^2}{2\mu} = -\frac{\hbar^2}{2\mu} \nabla^2\,,\\
  \nabla^2 &=& \left( %
  \frac{1}{r}\frac{\partial}{\partial r} +
  \frac{\partial^2}{\partial r^2} +
  \frac{1}{r^2}\frac{\partial^2}{\partial\theta^2}
  \right)\,,
\end{eqnarray}
\end{subequations}
with the time-dependent Dirichlet boundary conditions:
\begin{eqnarray}
  \psi(\lambda(t)\gamma(\theta), \theta) = 0\,.
\end{eqnarray}

More precisely, the Hamiltonian is given by an operator which coincides, time by time, with the kinetic term in the domain corresponding to the box, and zero elsewhere. For the sake of simplicity we will avoid introduction of such a new symbol.

According to our treatment in \cite{ref:DiMartino} and \cite{ref:Anza2015}, we make a unitary transformation which essentially maps the original domain delimited by the boundary given by \eqref{eq:OriginalBoundary} into another domain with the same shape but different diameter:
\begin{equation}\label{eq:RescaledBoundary}
  r = \gamma(\theta)\,,\qquad \theta \in [0, 2\pi]\,.
\end{equation}
We call the first domain (the time-dependent one) ${\cal D}_\lambda$ and this second one (the static one) ${\cal D}_1$. The relevant unitary operator acts as follows:
\begin{equation}\label{eq:UnitaryMapping}
  \phi(r, \theta) = (U_\lambda \psi)(r, \theta) = \lambda \psi(r \lambda,
  \theta)\,.
\end{equation} 
It maps quantum states defined in $\dom_\lambda$ into states defined in $\dom_1$.

In the new picture, the operator which generates the dynamics of the particle is given by
\begin{eqnarray}\label{eq:EffectiveHamiltonian}
\nonumber H_{\mathrm{eff}} &=& U_\lambda H U_\lambda^\dag +
\ii\hbar\dot{U}_\lambda U_\lambda^\dag\\
&=& -\frac{\hbar^2}{2\mu \lambda^2} \nabla^2 + \ii\hbar
\frac{\dot{\lambda}}{\lambda} \left(1 + \, r
\frac{\partial}{\partial
  r}\right) \,,
\end{eqnarray}
where one has to take into account the fact that, since $\lambda$ depends on $t$, the unitary operator $U_\lambda$ depends on time as well. The Hamiltonian in \eqref{eq:EffectiveHamiltonian} describes a particle with varying mass (because of the factor $\lambda^{-2}$ in the kinetic energy) in the presence of a time-dependent \lq potential\rq\,, i.e., the term $\ii\hbar\dot{\lambda}/\lambda(1+r\partial_r)$, which from now on we will call the \lq dilation potential\rq\, or \lq dilation term\rq.

It is remarkable to note that the dilation potential involves only the radial coordinate. This is a consequence of the pantographic nature of the box movement, which implies radial dilation and then only radial motion of the walls of the box. Of course, this simple fact does not guarantee conservation of the angular momentum. Indeed, generally speaking, the commutator between two operators depends on the domain in which the two operators are acting. Therefore, commutation between the dilation potential and the angular momentum will depend also on the shape of the box.

It is also worth noting that in the more general case of changes with deformation (i.e., non pantographic) the effective Hamiltonian in the static domain turns out to be much more complicated than the one in \eqref{eq:EffectiveHamiltonian} (the complete expression is reported in Ref.~\cite{ref:Anza2015}).

\subsection{One-dimensional and three-dimensional cases}

The one-dimensional and three-dimensional counterparts of our problem are treated in a very similar way. In the one-dimensional case, the domain is expressed as $\dom_\lambda=[-\lambda(t)l/2, \lambda(t)l/2]$ and will be mapped into $\dom_1=[-l/2, l/2]$, the boundary conditions are $\psi(\pm\lambda(t)/2) = 0$ and will be mapped into $\phi(\pm l/2) = 0$, and the generator of the time evolution will change as follows:
\begin{equation}
  H=-\frac{\hbar^2}{2\mu}\frac{\partial^2}{\partial x^2}%
  \;\;\;\;\Longrightarrow \;\;\;\;%
  H_{\mathrm{eff}} = -\frac{\hbar^2}{2\mu \lambda^2} \frac{\partial^2}{\partial x^2} + \ii\hbar\frac{\dot{\lambda}}{\lambda} \left(\frac{1}{2} + \, x \frac{\partial}{\partial
  x}\right)\,,
\end{equation}
where the last differential operator can be put in the form $r\partial_r$, with $r=|x|$.

In the three-dimensional case, the domain is expressed as $\dom_\lambda=\{(r,\theta,\varphi)| r\le \lambda(t)\gamma(\theta,\varphi), \theta\in [0,2\pi], \varphi\in [0,\pi]\}$ and will be mapped into $\dom_1$ obtained for $\lambda=1$, the boundary conditions are $\psi(\lambda(t)\gamma(\theta,\varphi),\theta,\varphi) = 0$ $\forall \theta, \varphi$ and will be mapped into $\phi(\gamma(\theta,\varphi),\theta,\varphi) = 0$, and the generator of the time evolution will change as follows:
\begin{equation}
  H=-\frac{\hbar^2}{2\mu}\nabla^2%
  \;\;\;\;\Longrightarrow \;\;\;\;%
  H_{\mathrm{eff}} = -\frac{\hbar^2}{2\mu \lambda^2} \nabla^2 + \ii\hbar\frac{\dot{\lambda}}{\lambda} \left(\frac{3}{2} + \, r \frac{\partial}{\partial
  r}\right)\,.
\end{equation}

This clarifies how the further results, which will be explicitly derived for the two-dimensional case, are essentially valid in 1D and 3D.

\section{Pantographic perturbations}\label{sec:Resonances}

Once the problem of one particle in a 2D box with moving walls is transformed into the problem of a particle in a static box, we get the Hamiltonian in \eqref{eq:EffectiveHamiltonian}.

Under the assumption that the dilation parameter $\lambda$ is a smooth, non-rapidly varying and close-to-unity function, the quantity $\dot{\lambda}/\lambda$ is small and the dilation potential can be treated as a perturbation. This is in perfect agreement with considerations by Kato on the way to treat small modifications of the domain in which a particle can move \cite{ref:Kato}. Such an approach has already been exploited in Ref.~\cite{ref:Anza2015} to treat changes of the domain with deformation. It is appropriate to mention here that in \cite{ref:Anza2015} we assume knowledge of the dynamics in the pantographic case --- the relevant Hamiltonian describing pantographic changes is considered as the unperturbed one --- and then treat the terms coming from deformation as a perturbation. Nevertheless, knowledge of the dynamics in the pantographic case is quite limited, and exact dynamics is known only for the case of uniformly moving domain. In this paper, instead, we are attempting to extend our knowledge of the dynamics in the pantographic case.

In the new picture, here addressed as the \lq Schr\"odinger picture\rq, generated by making the unitary mapping from $\dom_\lambda$ to $\dom_1$, we have a static domain and a time-dependent Hamiltonian (see \eqref{eq:EffectiveHamiltonian}) of the following form:
\begin{equation}
  H_\mathrm{S}(t) = \lambda^{-2}\, H_\mathrm{0} + \dot\lambda\,\lambda^{-1}\,
  V\,.
\end{equation}

By performing the passage to the interaction picture through the unitary operator generated by the \lq unperturbed\rq\, Hamiltonian $\lambda^{-2}\, H_\mathrm{0}$, $U_\mathrm{0}(t) = \e^{-\ii
\hbar^{-1} \int_0^t [\lambda(s)]^{-2}\mathrm{d}s\, H_\mathrm{0}}$, we obtain the new generator of the time evolution, $H_\mathrm{I}(t) = \dot\lambda\,\lambda^{-1} \, U_\mathrm{0}^\dag(t) \, V \, U_\mathrm{0}(t)$.

We now express $\lambda$ as follows:
\begin{equation}
\lambda = 1+\epsilon f(t)\,,
\end{equation}
where $\epsilon$ is a dimensionless small parameter, $f(t)$ is a smooth, non-rapidly varying and bounded function, and the operators $H_\mathrm{0}$ and $V$ are time-independent. We also introduce the series expansions of $\lambda^{-2}$ and $\dot\lambda\,\lambda^{-1}$:
\begin{eqnarray}
\lambda^{-2} &=& \sum_{\mathrm{k}}
A_\mathrm{k}\,\epsilon^\mathrm{k} = 1 - 2\epsilon f(t) +
...\,, \\
\dot\lambda\,\lambda^{-1} &=& \sum_{\mathrm{k}}
B_\mathrm{k}\,\epsilon^\mathrm{k} = \epsilon \dot{f}(t) + ...\,,
\end{eqnarray}
then getting the following Hamiltonian corrected to the first order in $\epsilon$:
\begin{equation}
  H_\mathrm{S}(t) = (1 - 2\epsilon f(t))\, H_\mathrm{0} + \epsilon \dot{f}(t)\,
  V\,.
\end{equation}

After performing the relevant spectral decompositions,
\begin{equation}
  H_\mathrm{0} = \sum_\alpha E_\alpha
  \Ket{E_\alpha}\Bra{E_\alpha}\,,
\end{equation}
and applying the standard perturbation treatment, we can explicitly write down the evolved state corrected up to the first order in the parameter $\epsilon$:
\begin{equation}
  \Ket{\psi(t)} = \left[\e^{-\frac{\ii}{\hbar} \int_0^t [1-2\epsilon f(s)]\mathrm{d}s\, H_\mathrm{0}}  +
  \epsilon \, \e^{-\frac{\ii}{\hbar}  H_\mathrm{0} t} \sum_{\alpha\beta}
  V_{\alpha\beta} \, \int_0^t \dot{f}(u) \, e^{\frac{\ii}{\hbar}
  (E_\mathrm{\alpha} - E_\mathrm{\beta}) u} \mathrm{d}u
  \Ket{E_\mathrm{\alpha}}\Bra{E_\mathrm{\beta}}\right] \Ket{\psi(0)}\,,
  \label{eq:PerturbativeEvolution}
\end{equation}
where $V_{\alpha\beta} = \Bra{E_\alpha} V \Ket{E_\beta}$.

On the basis of \eqref{eq:PerturbativeEvolution} one can immediately argue that the movement of the walls induces quantum transitions, and two factors determine which transitions are allowed. On the one hand the coupling through the dilation potential, which means that the matrix element $V_{\alpha\beta}$ must be nonzero. On the other hand, in order to have a finite transition probability from the initial state to another at long time, the following condition must be satisfied:
\begin{equation}\label{eq:FourierTransform}
\int_0^t \dot{f}(u) \, e^{\frac{\ii}{\hbar}
  (E_\alpha - E_\beta) u} \mathrm{d}u \not= 0\,,\qquad t\gg
  \frac{\hbar}{|E_\alpha - E_\beta|}\,.
\end{equation}

Since the integral in \eqref{eq:FourierTransform} in the limit $t\rightarrow \infty$ essentially approaches the Fourier transform of $\dot{f}(t)$ for $\omega = -(E_\alpha - E_\beta)/\hbar$, one can assert that in order to have nonzero transitions between two states the Fourier transform of the radial velocity of the walls should be nonzero for the relevant frequency. On this basis, it is natural to talk about resonances and turns out to be important singling out the presence of resonant sinusoidal components in the perturbation (i.e., the dilation potential). Of course, as usual, non resonant components in the Fourier expansion of the perturbation can induce fluctuations determining small transitions at small time, but such transitions disappear after a sufficiently long time.

As a relevant physical situation, one can consider the case where the $\lambda$ is a periodic function, and still smooth and close-to-unity. This means that $f(t)$ is smooth and periodic, and so is $\dot{f}(t)$, easily leading to the fact that, at first order, only transitions associated to the frequency of $\dot{f}$ or its multiples are allowed.

\section{The \lq Breathing\rq\, Circle}\label{sec:OscillatingCircle}

In order to better fix the previous ideas we analyze a very special case, that we call the \lq breathing\rq\, circle. In other words, we consider a particle moving inside a two-dimensional circular box whose radius is time-dependent: $R(t) = R_0 (1+\epsilon\sin\omega t)$, so that $\gamma(\theta)=R_0$ and $f(t)=\sin\omega t$. After mapping the original problem into the problem of a particle confined in a circular box of radius $R_0$ and expanding the relevant Hamiltonian with respect to $\epsilon$ up to the first order, one obtains:
\begin{equation}\label{eq:EffCircHamiltonian}
H_{\mathrm{eff}} = -\frac{\hbar^2}{2\mu} (1-2\epsilon\sin\omega t)
\nabla^2 + \epsilon\, \omega\, \cos\omega t \, \ii\hbar \, \left(1
+ \, r \frac{\partial}{\partial
  r}\right) \,.
\end{equation}

According to the analysis in Ref.~\cite{ref:Robinett1996}, the eigenvalues and eigenfunctions of a \lq free\rq\, particle (hence governed by $H_\mathrm{0}=-\hbar^2/(2\mu)\nabla^2$) in a 2D circular box are given by the following expressions:
\begin{eqnarray}
  E_{mn} &=& \frac{\hbar^2}{2 \mu r_0^2} a_{mn}^2\,,\\
  \chi_{mn} &=& (2\pi)^{-1/2} \aleph_{mn} J_{m}(k_{mn} r ) \times e^{\ii m \theta}\,, \qquad m \in
  \mathbb{Z}\,,
\end{eqnarray}
where $J_m(x)$ is the Bessel function of $m$-th order, $a_{mn}$ is
the $n$-th zero of $J_m$, and
\begin{eqnarray}
  && k_{mn}^2 = \frac{2 \mu E_{mn}}{\hbar^2} = \frac{(a_{mn})^2}{r_0^2} \; \Rightarrow \; k_{mn} = \frac{|a_{mn}|}{r_0} \, ,\\
  && \aleph_{mn} = \left(\int_0^{r_0} r J_m (k_{mn}r)^2
  \mathrm{d}r\,\right)^{-1/2}\,.
\end{eqnarray}

We have already mentioned the fact that the operator $1+r\partial_r$ does not involve the angular variable but only the radial one. Moreover, because of the specific shape of the box, in this case the angular momentum of the system commutes with such an operator. This fact, considered also the commutation between kinetic energy and angular momentum, implies that the angular momentum is conserved through all the evolution. This result is indeed quite intuitive since the shape of the box is circular at any time and the \lq kicks\rq\, that the particle can receive from the walls of the box are only directed along the radius of the circle, hence not producing any \lq torque\rq. On the contrary, because of such kicks, the energy of the particle can change, which corresponds to the possibility of having transitions between states with the same angular momentum but different energies.

In connection with the perturbation treatment, we can say that the matrix elements of the dilation potential are vanishing when different values of the angular momentum quantum number are involved: $V_{mnm'n'}=\Bra{\chi_{mn}}V\Ket{\chi_{m'n'}}=\delta_{m m'} \Bra{\chi_{mn}}V\Ket{\chi_{mn'}}$.

\section{Discussion}\label{sec:Conclusion}

The analysis developed in previous works, such as Ref.~\cite{ref:DiMartino} and Ref.~\cite{ref:Anza2015}, shows that when the problem of a system with moving boundaries is mapped into a problem with static boundaries the Hamiltonian inducing the evolution in the new picture becomes time-dependent. Even  the very kinetic energy contains a time-dependent factor, so that the particle appears as a particle with changing mass. Moreover, a new term - the dilation term, which is time-dependent as well - is added. Such new term can induce transitions between the eigenstates of the unperturbed Hamiltonian, which in our case is simply given by the kinetic energy of a particle with a varying mass. Which transitions are induced depends on the way the boundary moves, and in particular from the Fourier transform of the velocity of the walls. As a very special case, we have considered the case where the walls oscillate at a precise frequency, so that transitions between eigenstates of the \lq free\rq\, Hamiltonian can be induced when the boundary oscillations are properly \lq tuned\rq .

We emphasize the fact that our analysis demonstrates with an appropriate mathematical treatment the intuitive idea that a boundary oscillating at a given frequency can induce transitions, when such oscillations are tuned to a transition frequency of the system. Then, in some sense, an oscillating boundary acts on the system like a suitable oscillating field. By this way, it is worth mentioning that, because of the choice of considering only pantographic changes of the domain, the dilation potential\, turns out to be a radial potential ($1+r\partial_r$) and then, with an appropriate geometrical nature of the domain, it preserves angular momentum, as it happens in the case of a circular box.

As a final remark, we point out that if the particle in the original time-dependent domain is subjected, not only to the potential describing the box contour, but also to another potential, of course such a term will be kept in the new picture with a suitable scaling (see Ref.~\cite{ref:DiMartino}). In our case, for the sake of simplicity we have not considered such a situation. Nevertheless, in the presence of such additional potential the approach is essentially the same as before, just with some mathematical complications. In fact, in the static domain the particle would be described as a time varying particle subjected to a time-dependent potential (resulting as a time-dependent scaling of the original potential) and to the dilation term (resulting from the change of picture). Therefore, in the limit of small and smooth movements of the walls of the box, the last term will be treated as a perturbation to the time evolution induced by the first two terms.

\end{document}